\documentclass[a4paper,twocolumn,pra,showpacs,showkeywords,aps,nofootinbib]{revtex4}
\usepackage{amsmath,amsfonts,amssymb}
\usepackage{graphicx}
\usepackage{mathbbol}
\newcommand{\beq}{\begin{equation}}
\newcommand{\eeq}{\end{equation}}

\newcommand{\beqa}{\begin{eqnarray}}
\newcommand{\eeqa}{\end{eqnarray}}
\def\ra{\rangle}
\def\la{\langle}

\begin{document}
\title{Fast atomic transport without vibrational heating}

\author{E. Torrontegui$^{1}$}
\author{S. Ib\'a\~nez$^{1}$} 
\author{Xi Chen$^{1,2}$} 
\author{A. Ruschhaupt$^{3}$}
\author{D. Gu\'ery-Odelin$^{4}$}
\author{J. G. Muga$^{1,4,5}$}

\affiliation{$^{1}$Departamento de Qu\'{\i}mica F\'{\i}sica, Universidad del
Pa\'{\i}s Vasco, Apartado 644, 48080 Bilbao, Spain}

\affiliation{$^{2}$Department of Physics, Shanghai University,
200444 Shanghai, P. R. China}

\affiliation{$^{3}$Institut f\"ur Theoretische Physik, Leibniz
Universit\"{a}t Hannover, Appelstra$\beta$e 2, 30167 Hannover,
Germany}

\affiliation{$^{4}$Laboratoire Collisions Agr\'egats R\'eactivit\'e, CNRS UMR 5589, IRSAMC, Universit\'e Paul Sabatier, 118 Route de Narbonne, 31062 Toulouse CEDEX 4, France}

\affiliation{$^{5}$Max Planck Institute for the Physics of Complex Systems, 
N\"othnitzer Str. 38, 01187 Dresden, Germany}
\begin{abstract}
We use the dynamical invariants associated with the Hamiltonian of an atom in a one dimensional moving trap to inverse engineer the trap motion 
and perform fast atomic transport without final vibrational heating.  
The atom is driven non-adiabatically through a shortcut to the result of adiabatic, slow trap motion.  For harmonic potentials this only requires designing appropriate trap trajectories, whereas perfect transport in anharmonic traps may be achieved by  applying an extra field to compensate the forces in the rest frame of the trap. 
The results can be extended to atom stopping or launching. 
The limitations due to geometrical constraints, energies and accelerations involved are analyzed, as well as the relation to previous approaches (based on classical trajectories or ``fast-forward'' and ``bang-bang'' methods) which can be   
integrated in the invariant-based framework.   
\end{abstract}
\maketitle
\section{Introduction}
A key element to attain an exhaustive control of states and dynamics
of cold atoms and ions is their efficient 
transport by moving the confining trap. 
In spite of the broad span of conditions, heating mechanisms, transport distances (from microns to tens of centimeters), transport times, and  accelerations involved, there are some common elements and objectives that allow for a rather generic theoretical treatment
as the one presented in this paper.
Transport should ideally be lossless, fast, and 
lead to a final state as close as possible (``faithful'') to the initial one, up to global phase factors, in the frame of the transporting trap. 
The later requirement is characterized in the most demanding applications as a high-fidelity condition, or more generally as a no-heating or at least minimal heating condition; equivalently, by the absence or minimization
of vibrational excitations at the end of the transport. 
Note that reaching a faithful final state is not incompatible with some transient excitation in the instantaneous basis at intermediate times, 
i.e., the process does not have to be slow (in the usual quantum mechanical jargon, ``adiabatic''), although    
slowness is certainly a simple way to avoid heating, at least for ideal conditions.

Efficient atom transport is a major goal for many applications
such as quantum information processing in multiplexed trap arrays \cite{Leibfried2002,ions} or quantum registers \cite{MeschNature}; 
controlled translation from the
production (cooling) chamber to interaction or manipulation zones \cite{Ketterle2002,Cornell2003,Pritchard};  
accurate control of interaction times and locations, e.g. in 
cavity QED experiments \cite{detector}, quantum gates \cite{Calarco2000} or metrology \cite{Maleki}; and velocity control to launch 
\cite{Meschede},  or stop atoms \cite{catcher1,catcher2}. 

Different approaches have been implemented. 
Neutral atoms have been transported as thermal atomic clouds \cite{HanschPRL2001,Pritchard}, 
condensates \cite{HanschNature2001}, or individually \cite{ocnsna,ocn}, 
using magnetic, or optical traps.
The magnetic traps 
can be translated by moving the coils mechanically \cite{Cornell2003}, by  time-varying currents in a lithographic conductor pattern  \cite{HanschPRA2001}, or on a conveyor belt with a chain of permanent magnets \cite{Lahaye}. Optical traps can be used as optical tweezers whose focal point is translated by moving mechanically lenses \cite{Ketterle2002,David}, or by traveling lattices (conveyor belts) 
made with two counterpropagating beams slightly detuned one respect to each other \cite{ocnsna,ocn,Denschlag}. There are also mixed magneto-optical 
approaches \cite{Pritchard}. For ions, controlled time dependent voltages
have been used in linear-trap based frequency standards \cite{Maleki}
and more recently in quantum information applications using multisegmented Paul traps \cite{SK}, an array of Penning traps \cite{Penning}, and also in 2D configurations \cite{Wineland}.
 

As said before, an obvious solution, at least in principle, to avoid spilling or heating of the atoms 
is to perform a sufficiently slow (adiabatic) transport. 
For some applications, however, this takes too long.   
In particular, since transport could occupy most of the operation time 
of realistic quantum information algorithms, ``shuttling times'' need
to be minimized \cite{ions,SK}. In addition, long times may be counterproductive in practice and induce overheating from coils or fluctuating fields and decoherence. There are in summary good reasons to 
reduce the transport time, and indeed    
several theory and experimental works have studied ways to make fast transport  
also faithful \cite{David,Calarco2009,Nakamura2010,Shan}.

  
Invariant-based inverse engineering is ideally suited to this end. The main aim of this paper is to set the basic invariant-based inverse engineering transport theory, analog to the one developed recently for trap expansions \cite{Chen}. We shall also show that previous approaches for  efficient transport \cite{Calarco2009,Nakamura2010} and some generalizations 
are embraced by it, and  point out the 
potential limitations of the method.
In Sec. \ref{sec2} we shall provide the main concepts and
formulae of the time dependent quadratic-in-momentum invariants  
relevant for transport problems. The two main reference cases are ({\it{i}})  rigid harmonic oscillator transport and ({\it{ii}}) transport on an arbitrary trap with force compensation. In Sec. \ref{ie} we explain and apply the inversion technique; this is compared in Sec. \ref{bang-bang} 
with an alternative ``bang-bang'' approach based on time segments of constant acceleration.  Section \ref{tee}
deals with practical limitations and Sec. \ref{dis}
discusses the results and draws the conclusions.              
\section{Dynamical Invariants\label{sec2}}
In a seminal paper Lewis and Riesenfeld derived a simple relation between the solutions of the Schr\"odinger equation of a system with time-dependent Hamiltonian and the eigenstates of the corresponding invariants \cite{LR}. They paid special attention to the time-dependent harmonic oscillator and its invariants quadratic in position and momentum,
related to earlier work by Ermakov on the classical oscillator \cite{Ermakov}. From a classical physics point of view, Lewis and Leach found the general form of the Hamiltonian compatible with invariants quadratic in momentum \cite{LL}, including non harmonic potentials. This is the result that interests us here, together with the corresponding quantum formulation by Dhara and Lawande \cite{DL}. In this Section we shall state the main concepts and equations and apply them to standard transport problems.    

A 1D Hamiltonian with an invariant which is quadratic in momentum must have the form $H=p^2/2m+V(q,t)$,\footnote{Following the usual practice, $q$ and $p$ will denote operators or numbers, and the context should clarify their meaning.}
with \cite{LL,DL}   
\beq
\label{Vinv}
V(q,t)=
-F(t)q+\frac{m}{2}\omega^2(t)q^2+\frac{1}{\rho(t)^2}U\left[\frac{q-\alpha(t)}{\rho(t)}\right].  
\eeq
$\rho$, $\alpha$, $\omega$, and $F$ are arbitrary functions of time 
that 
satisfy the auxiliary equations 
\beqa
\ddot{\rho}+\omega^2(t)\rho=\frac{\omega_0^2}{\rho^3},
\label{Erma}
\\
\ddot{\alpha}+\omega^2(t)\alpha=F(t)/m,  
\label{alphaeq}
\eeqa
with $\omega_0$ constant. Their physical interpretation in the context of transport is detailed below. 
The dynamical invariant, up to a constant factor,
is given by 
\beqa
\label{inva}
I&=&\frac{1}{2m}[\rho(p-m\dot{\alpha})-m\dot{\rho}(q-\alpha)]^2
\nonumber\\
&+&\frac{1}{2}m\omega_0^2\left(\frac{q-\alpha}{\rho}\right)^2
+U\left(\frac{q-\alpha}{\rho}\right)
\eeqa
and verifies
\beq
\frac{ {\rm d} I  }{  {\rm d} t }\equiv
\frac{\partial I(t)}{\partial t} +\frac{1}{i\hbar} [I(t),H(t)]=0, 
\eeq
so that $\frac{d}{dt}\la \psi(t)|I(t)|\psi(t)\ra=0$ for any wave function $\psi(t)$ that evolves with $H$.  
$\psi(t)$ may be expanded in terms of constant coefficients $c_n$ and eigenvectors $\psi_n$ of $I$, 
\beqa
\psi(q,t)&=&\sum_n c_n e^{i\alpha_n} \psi_n(q,t),  
\\
I(t)\psi_n(q,t)&=&\lambda_n\psi_n(q,t),
\eeqa
where the $\lambda_n$ are time-independent eigenvalues.      
%
We shall generally deal with $\psi_n$ normalized to one, but continuum, delta-normalized states are also possible.   
The phases $\alpha_n$ satisfy \cite{LR,DL}
\beqa
\hbar\, \frac{d \alpha_n}{dt}&=&\bigg\la\psi_n\bigg|i\hbar\frac{\partial}{\partial t}
-H\bigg|\psi_n\bigg\ra,
\\
\alpha_n&=&-\frac{1}{\hbar}\int_0^t  {\rm d} t'\left(\frac{\lambda_n}{\rho^2}+
\frac{m(\dot{\alpha}\rho-\alpha\dot{\rho})^2}{2\rho^2}\right).
\eeqa
The $\psi_n$ are in practice obtained easily as \cite{DL}  
\beq\label{psin}
\psi_n (q,t)=e^{\frac{im}{\hbar}\left[\dot{\rho} q^2/2\rho+(\dot{\alpha}\rho-\alpha\dot{\rho})q/\rho\right]}\frac{1}{\rho^{1/2}}\phi_n\bigg(\underbrace{\frac{q-\alpha}{\rho}}_{=:\sigma}\bigg)
\eeq
from the solutions $\phi_n(\sigma)$ (normalized in $\sigma$-space) of the auxiliary stationary Schr\"odinger equation
\beq
\left[-\frac{\hbar^2}{2m}\frac{\partial^2}{\partial\sigma^2}+\frac{1}{2}m\omega_0^2\sigma^2+U(\sigma)\right]\phi_n=\lambda_n\phi_n.
\label{last}
\eeq
%
Whereas trap expansions and contractions imply 
a time dependent
$\rho$ function \cite{Chen},
a large family of transport problems may be described
by taking 
\beq\label{condtran}
\rho(t)=1,\; \omega^2(t)=\omega_0^2
\eeq
so the auxiliary Eq. (\ref{Erma}) 
plays no role and only Eq. (\ref{alphaeq}) is relevant.
Except in the final discussion we shall  
assume that the conditions (\ref{condtran}) hold from now on, 
and consider in detail two main reference cases. 
\subsection{Main cases}

({\it{i}}) {\it{Rigid harmonic oscillator driven by the
``transport function''}} $q_0(t)$. (Hereafter ``harmonic transport'' for short.) Suppose that a harmonic trap is moved from $q_0(0)$ at time $t=0$ to $d=q_0(t_f)$ at a time $t_f$.
In Eq. (\ref{Vinv}) this case corresponds to 
\beq
F=m\omega_0^2 q_0(t),\; \omega(t)=\omega_0,\;U=0.
\eeq
Adding to $V$ the irrelevant   
time dependent global term $m\omega_0^2 q_0^2/2$, which produces no force,  the trap 
potential can be written as a moving harmonic oscillator $m\omega_0^2[q-q_0(t)]^2/2$, 
\beq
\label{hamiht}
H=p^2/2m+m\omega_0^2[q-q_0(t)]^2/2, 
\eeq
%
and $\alpha$ may be identified with a classical trajectory $q_c$ since Eq. (\ref{alphaeq}) becomes 
\beq
\label{classical}
\ddot{q}_c+\omega_0^2(q_c-q_0)=0.
\eeq
The invariants and transport modes will depend on it. 
In this case $\lambda_n=E_n=(n+1/2)\hbar \omega_0$, and
the transport mode $e^{i\alpha_n}\psi_n$ takes a physically 
transparent form,   
\beq\label{psin}
e^{i\alpha_n}\psi_n=e^{-\frac{i}{\hbar}\left[E_n t+\int_0^{t} \frac{m\dot{q}_c^2}{2} dt'\right]} 
e^{im\dot{q}_c q/\hbar}\phi_n({q-q_c}).
\eeq
Efficient transport will be engineered in the following section by designing first an appropriate classical trajectory $q_c(t)$, from which the trap motion trajectory $q_0(t)$ is deduced via Eq. (\ref{classical}). 
 
A variant of this case is vertical transport with a gravity force, so that  
$F=m\omega_0^2q_0-mg$ and Eq. (\ref{classical}) is modified to  
\beq
\ddot{q}_c+\omega_0^2(q_c-q_0)=-g.
\eeq  

({\it{ii}}) {\it{Arbitrary-trap driven transport with compensating force}}. (Hereafter ``compensating force approach" for short.) 
Now, in Eq. (\ref{Vinv}) 
\beqa
\omega&=&\omega_0=0,
\\
F&=&m\ddot{q}_0.
\label{compen}
\eeqa
In this case the trap potential $U[q-q_0(t)]$ is arbitrary (in particular it could be harmonic), and it is rigidly displaced along $q_0(t)$, so $\alpha$ in Eq. (\ref{alphaeq}) may be now identified with the transport function $q_0$. In addition to $U$, 
there is a  time dependent linear potential term $-mq\ddot{q}_0$ in $H$, 
\beq
\label{hamii}
H=p^2/2m-mq\ddot{q}_0+U(q-q_0).
\eeq
The corresponding force compensates exactly the inertial force due to the trap motion in the rest frame of the trap, so that the wave function in that frame is not modified up to a time dependent global phase factor, see the Appendix \ref{B}.  
This Hamiltonian has been proposed by Mashuda and Nakamura following a very different route, using a ``fast-forward'' scaling technique  \cite{Nakamura2010}.  
\section{Inverse engineering method\label{ie}}
The Lewis-Riesenfeld theory of invariants has been considered before in harmonic-oscillator driven transport in the direct (rather than inverse)  way, by setting the transport function $q_0$ and analyzing the final heating, in particular in adiabatic or quasi-adiabatic regimes \cite{ions}.
We shall use it instead as the basis for an inverse engineering approach, 
including also non-harmonic driving.
The two main cases discussed above require different 
implementations.  In both cases we shall assume that  $q_0$ is displaced from $0$ to $d$ in a time $t_f$. 
\subsection{Harmonic transport\label{htr}}   
In case ({\it{i}}) we may adopt as in \cite{Chen} an inverse engineering strategy by designing first the classical trajectory $q_c$ to assure 
that the transport modes coincide with the eigenvectors of the instantaneous Hamiltonian at initial and final times. This amounts to impose the commutativity of $I(t)$ and $H(t)$ at $t=0$ and $t=t_f$, which can be achieved by setting, see Eq. (\ref{psin}), the following boundary conditions,
\beqa
\label{con0}
q_c(0)&=&q_0(0)=0;\; \dot{q}_c(0)=0;\; \ddot{q}_c(0)=0,
\\
\label{contf}
q_c(t_f)&=&q_0(t_f)=d;\; \dot{q}_c(t_f)=0;\; \ddot{q}_c(t_f)=0,
\eeqa
the last conditions in each line being determined by consistency 
with Eq. (\ref{classical}). 
$q_c(t)$ is then interpolated by assuming some flexible functional form, such as a polynomial, $q_c(t)=\sum_{n=0}^5 \beta_n t^n$, where the $\beta_n$ are found   
by solving the system of equations established by the boundary conditions. 
The resulting $q_c$ depends on time only trough the ratio $s=t/t_f$, and is directly proportional
to $d$,   
\beq\label{pol}
q_c(t)/d=10 s^3-15 s^4+6 s^5.
\eeq
Once $q_c$ is fixed we get the trap trajectory from Eq. (\ref{classical}), 
\beq
q_0(t)=\ddot{q}_c(t)/\omega_0^2+q_c(t). 
\eeq
This procedure is equivalent to the 
one followed by Murphy et al. \cite{Calarco2009}, who used  
a Fourier sum as interpolating function for $q_c$. 
 
For short times $t_f$, see Fig. 1,  the corresponding trap trajectories $q_0(t)$ could exceed the interval 
$[0,d]$. For the polynomial ansatz (\ref{pol}) it occurs symmetrically at lower and upper edges of the interval for $t_f\le 2.505/\omega_0\approx 0.4$ $T_0$, where $T_0\equiv 2\pi/\omega_0$ is the oscillation period. 
This may or may not be a problem depending on the geometrical constraints of the experimental setting.  
\begin{figure}[t!]
    \begin{center}
        \includegraphics[width=8cm]{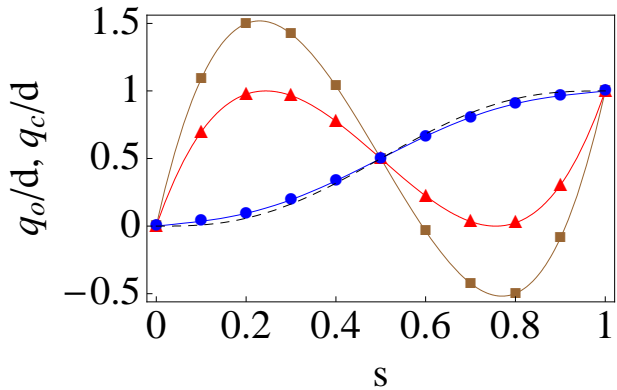}
    \end{center}
    \caption{(Color online) $q_0/d$ versus $s=t/t_f$ for  $t_f=12.57/\omega_0=2T_0$ (blue line with dots), $t_f=2.505/\omega_0$ (this is the critical value; red line with  triangles), and $t_f=2/\omega_0$ (brown line with squares). For all three cases $q_c/d$ is the dashed line, hardly distinguishable from $q_0/d$ for the slowest case, i.e., for $t_f=12.57/\omega_0$.}
\label{figq0}
\end{figure}

An interesting generalization is to consider boundary conditions for 
stopping atoms when their 
initial average velocity is known. Suppose that an atom gun or pulsed
valve sends atoms with a specific average velocity $v_0$, as in coil-gun experiments with paramagnetic atoms \cite{cg1,cg2}, or a Stark decelerator for polar molecules \cite{sd1,sd2,sd3}. 
\begin{figure}[t!]
    \begin{center}
        \includegraphics[width=8cm]{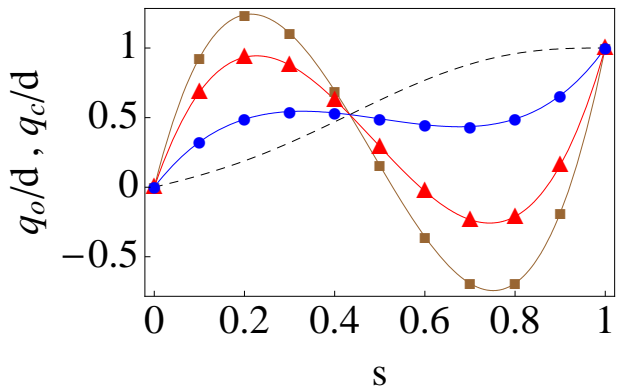}
    \end{center}
    \caption{(Color online) Three stopping
trajectories. $a=0.8$. 
$b=\omega_0 t_f=1.6$, $t_f/T_0=0.25$ (brown line with squares); 
$b=\omega_0 t_f=1.9$, $t_f/T_0=0.30$ (red line with triangles);  
$b=\omega_0 t_f=3.0$, $t_f/T_0=0.477$ (blue line with circles).  
They all share the same $q_c(s)/d$
(black dashed line) given by Eq. (\ref{qcs}).} 
\label{fig_stop}
\end{figure}
The valve opening time is controlled, so a traveling 
harmonic trap wrapping the atoms can be turned on at time $t=0$ and moved along some trajectory $q_0(t)$ 
to stop them  
at a fixed distance $d$ in a specified time $t_f$.    
The final conditions may still be given by Eq. (\ref{contf}), but a different set of initial conditions are to be imposed, 
\beq\label{inis}
q_c(0)=0,\; \dot{q}_c(0)=v_0,\; \ddot{q}_c(0)=0. 
\eeq
In this case the $n$-th initial transport mode at $t=0$ does not coincide with the $n$-th stationary eigenstate of $H(0)$ but with its moving version, $e^{imv_0q/\hbar}\phi_n(q)$.   
$\omega_0$ is in principle arbitrary, but it may be optimized taking into account the spatial width of the incoming state. Its value also has an impact on the domain of the trajectory as we shall see.  
The polynomial ansatz gives now
\beqa
q_c(t)&=&d [3(2-a)s^{5}-(15-8a)s^{4}
\nonumber
\\&-&2(-5+3a)s^{3}+as],\label{qcs}
\\ q_0(t)&=&d\bigg\{3(2-a)s^{5}-(15-8a)s^{4}
\nonumber\\
&+&\left[\frac{60(2-a)}{b^{2}}-2(-5+3a)\right]s^{3}-12\frac{(15-8a)}{b^{2}}s^{2}
\nonumber\\
&+&\left[-12\frac{(-5+3a)}{b^{2}}+a\right]s\bigg\},
\label{q0s}
\eeqa
where $s=t/t_f$, $a=v_0 t_f/d$ and $b=\omega_0 t_f=2\pi t_f/T_0$.
Some trajectories are shown in Fig. \ref{fig_stop}.    
The shaded regions in Fig. \ref{cor} 
correspond to the values of $a$ and $b/2\pi$ for which 
the trajectory $q_0(t)$ exceeds the domain $[0,d]$. 
Even though the details are now more complicated
than for the rest-to-rest case, two simple general rules
can be drawn: $q_0$ is never negative when
$t_f>T_0$ (the asymptotic threshold for large $a$ is at $b=6$), whereas if $a\ge2.513$ there is always some $t$
in $(0,t_f)$ for which $q_0(t)>d$.
%
\begin{figure}
\begin{center}
\includegraphics[width=4.cm]{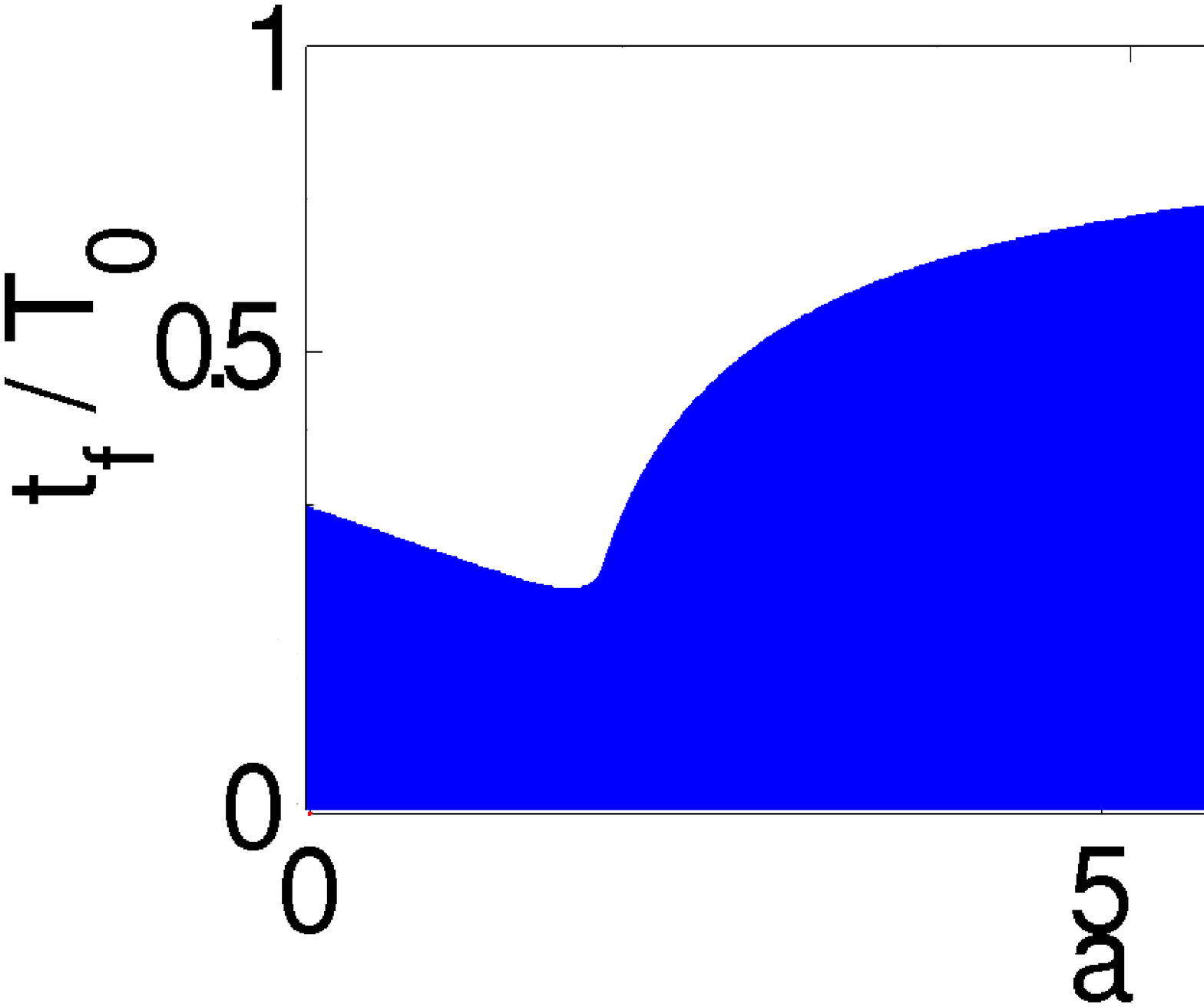}
\includegraphics[width=4.cm]{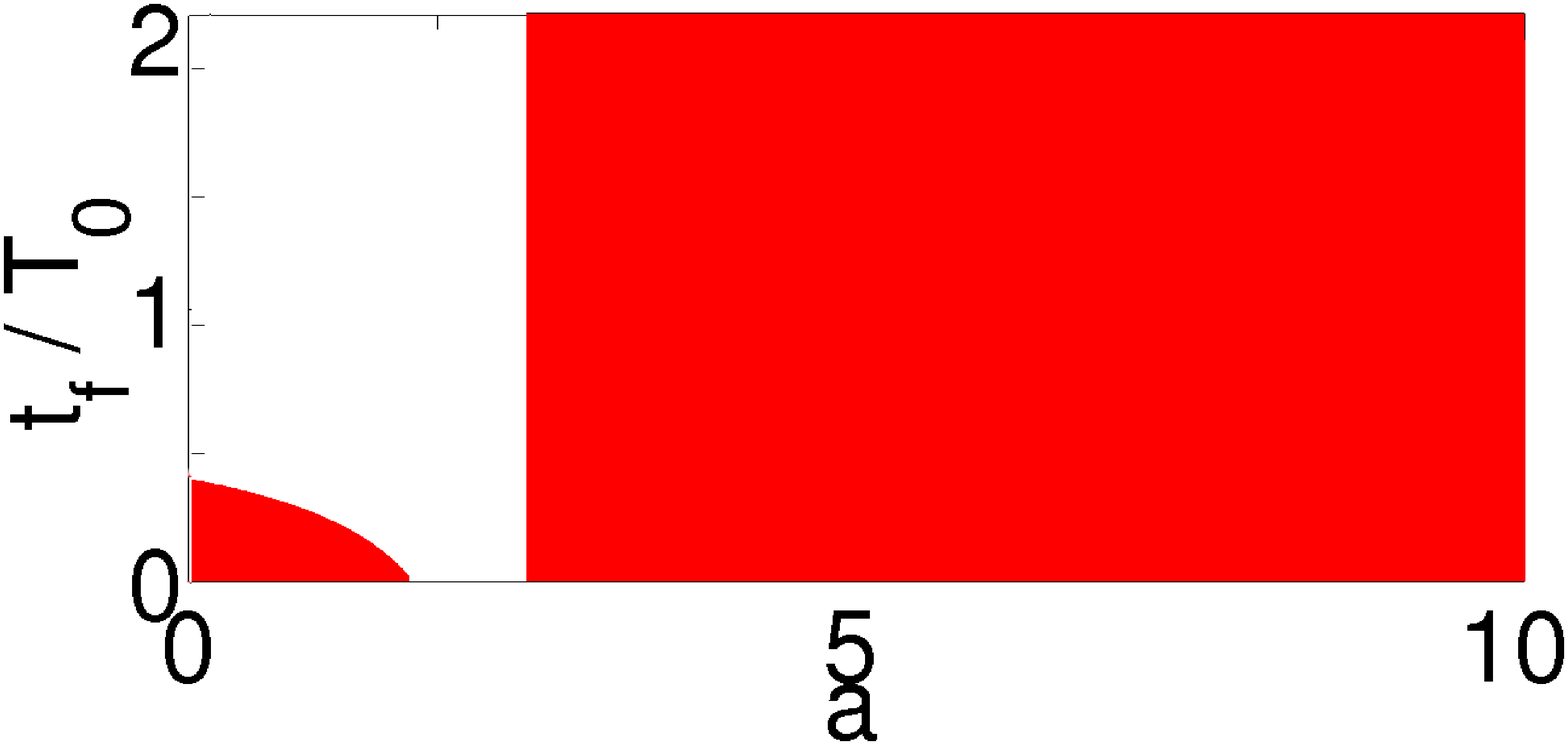}
\caption{(Color online) Stopping atoms: The shaded areas represent the values of $a=v_0t_f/d$ and $b/2\pi=t_f/T_0$ for which there is a $t$ in $(0, t_f)$ where $q_{0}(t)<0$ (left figure) 
or $q_{0}(t)>d$ (right figure). We have used the polynomial ansatz of 
Eqs. (\ref{qcs},\ref{q0s}).}
\label{cor}
\end{center}
\end{figure}
%
%
%

Launching or ``catapulting'' atoms at rest to end up with a chosen
velocity $v_0$ \cite{Meschede}
may be designed similarly
by setting Eq. (\ref{con0}) for the initial conditions, and the final boundary conditions as 
\beq
q_c(t_f)=d,\; \dot{q}_c(t_f)=v_0,\; \ddot{q}_c(t_f)=0. 
\eeq  
A major practical concern in all these applications should be to keep the 
harmonic approximation valid. This may require an analysis of the actual potential in each specific case and of the excitations taking place along the non-adiabatic transport process. Without performing such detailed analysis, the feasibility of the 
approach for a given transport objective set by the pair $d,t_f$ can be estimated rather simply by comparing lower excitation bounds provided in  Sec. \ref{tee} with the trap depth. 
\subsection{Compensating force approach}
In the compensating-force approach, case ({\it{ii}}),  we may proceed similarly but now the variable $\alpha$ of Eqs. (\ref{Vinv}) and (\ref{alphaeq}) is directly $q_0$ so, instead of fixing values of $q_c$ and its derivatives at the boundary times, we shall fix values of $q_0$ and its derivatives. For simplicity we shall consider only a rest-to-rest scenario, but the generalizations are straightforward. 
The compensating potential $-m\ddot{q}_0q$ in the
Hamiltonian (\ref{hamii}) should vanish  
before and after transport, since the trap remains at rest and $\ddot{q}_0=0$ for $t<0$ and $t>t_f$.  
To make the $\psi_n$ coincide, up to a global phase factor,  with the eigenstates of the Hamiltonian before and after transport, $H=p^2/2m+U(q-q_0)$,     
at $t=0$ and $t=t_f$, we impose 
\beqa
q_0(0)&=&\dot{q}_0(0)=0,
\nonumber\\
q_0(t_f)&=&d,\;\dot{q}_0(t_f)=0.
\label{condq0} 
\eeqa
%
We may as well impose $\ddot{q}_0=0$ as a boundary condition 
at $t=0$ and $t_f$, to have a continuous  
$\ddot{q}_0(t)$ but, at least formally, it is not strictly necessary. In practice   
several experiments have been designed with 
(approximate) discontinuities in the 
trap acceleration \cite{ocnsna,David}. We shall come back to this point below. 
Clearly the implementation of the compensating-force approach is subjected to different limitations from the ones applicable to the harmonic transport without compensation. The main problem now is not anharmonicity, 
which is included in the theory from the start by admitting an arbitrary $U$, but the feasibility of the compensating force term. 
According to the mean value theorem, see Sec. \ref{tee}, a lower bound for the 
maximum of the absolute value of $\ddot{q}_0$ is $2d/t_f^2$.  
\section{Bang-bang acceleration methods\label{bang-bang}}
It should be clear from section \ref{htr}, that 
there are infinitely many functions $q_0(t)$ which, 
for harmonic transport, lead to the ideal
boundary conditions. A somewhat extreme case, which has however been 
implemented experimentally because of its relative simplicity, is to   
combine time segments with a constant acceleration \cite{ocnsna,David}.  
The simplest trap trajectory of this type implies  
a constant positive acceleration $4d/t_f^2$ from $0$ to $t_f/2$, and a 
deceleration $-4d/t_f^2$ from $t_f/2$ to $t_f$. 
The resulting $q_0$ is formed by two     
parabolas matched at $t_f/2$, 
\beq\label{Davs}
q_{0}/d=\left\{
\begin{array}{ll}
{2s^2},& 0<s<1/2
\\
{4}(s-\frac{s^2}{2}-\frac{1}{4}),&
1/2<s<1
\end{array}
\right..
\eeq
The corresponding velocity $\dot{q}_0$ increases linearly from $0$ to 
$2d/t_f$ and then decreases from there to $0$.

The classical trajectory $q_c$ satisfying Eq. (\ref{classical}) with Eq. (\ref{Davs}), and the boundary conditions (\ref{con0}) at $t=0$ 
is given by 
\beq
\label{dife}
q_c-q_{0}=\left\{
\begin{array}{l}
\frac{-4d}{\omega_0^2t_{f}^{2}}(1-\cos\omega_0 t)
\\
\frac{4d}{\omega_0^2t_{f}^{2}}\bigg\{1+\cos\omega_0 t -2\cos\bigg[\omega_0\bigg(t-\frac{t_f}{2}\bigg)\bigg]\bigg\} 
\end{array}
\right.
\eeq
for the first and second time segments. 
From this result one can check that the boundary conditions at $t_f$ (\ref{contf}) are satisfied 
{\it only} for a discrete set of times $t_{f,N}=4\pi N/\omega_0$,  
with $N\in\mathbb{N}$, i.e., for multiples of two oscillation periods.  
For all other times this scheme will heat the atoms.  
A perturbation theory analysis 
shows that, even for the selected discrete times, the bang-bang 
method is less stable than the inverse invariant method with respect to 
an anharmonic perturbation of the transporting trap potential. The details are shown in Appendix \ref{pera}.       

A variant of this
method, using e.g. the trap trajectory (\ref{Davs})
and, in addition,
compensating forces as in case ({\it{ii}}), may be appealing
in practice because it is relatively simple to implement
the compensating force, at least approximately, as
a piecewise function.
%
%
%
%
%
%
%
%
%
\section{Transient energy excitations\label{tee}}
Whereas, ideally, non-adiabatic faithful transport can be performed for arbitrary transport distances and times, in practice the process could be limited, apart from the geometrical constraints discussed in the previous section, by the maximal transient excitation energies allowed to neglect the effect of anharmonicities of the actual potential in case ({\it{i}}), or by the difficulties to implement strong compensating forces in case ({\it{ii}}). 
We shall analyze these effects from the point of view of different bounds obtained for the 
average (potential) energy using the Euler-Lagrange equations, and for the instantaneous potential energy and acceleration
by means of the mean value theorem. 
%
%
%
%
%
\subsection{Quasi-optimal trajectories for harmonic  transport}
The instantaneous average energy for a harmonically-driven transport mode can be calculated from Eqs. (\ref{hamiht},\ref{psin}),    
\beqa
&&\langle \psi_n(t)|H(t)|\psi_n(t)\rangle = 
\nonumber \\
&&\hbar \omega_0\left(n+\frac{1}{2}\right)+\frac{m}{2}\dot{q}_c^2+\frac{1}{2}m
\omega_0^2(q_c-q_0)^2.  
\eeqa
Moreover the instantaneous average potential energy  is  $\la V(t)\ra=\frac{\hbar\omega_0}{2}\left(n+1/2\right)+E_P$.  
The first, ``internal'' contribution remains constant for each $n$,
and the second term  
has the simple form of a potential energy for a classical particle, $E_P\equiv\frac{1}{2}m\omega_0^2(q_c-q_0)^2$.  
Its time average, using the relation (\ref{classical})
between $q_0$ and $q_c$, takes the form 
%
%
%
\beq
\label{epot}
\overline{E_P}=\frac{m}{2t_f \omega_0^2}\int_0^{t_f} \ddot{q}_c^2 dt.
\eeq
We can use a generalized Euler-Lagrange equation $d^4q_c/dt^4=0$  
to minimize this integral subject to  
{\it four} boundary conditions \cite{Fo}, the ones for $q_c$ and $\dot{q}_c$ in Eqs. (\ref{con0},\ref{contf}).
This results in a ``quasi-optimal'' classical trajectory 
\beq
\label{fourbc}
q_c=d(3s^2-2s^3). 
\eeq
Whereas Eq. (\ref{fourbc}) does not satisfy the six boundary conditions
(\ref{con0},\ref{contf}), it provides in any case a lower bound for the time average of  $E_P$, as the set of functions satisfying the six conditions is smaller than the one satisfying four of them.    
Substituting Eq. (\ref{fourbc}) into Eq. (\ref{epot})
one gets the desired lower bound, 
\beq
\label{boundav}
{\overline{E_{P}}}\ge\frac{6md^2}{t_f^4 \omega_0^2}, 
\eeq
\begin{figure}[h!]
\begin{center}
\includegraphics[width=8cm]{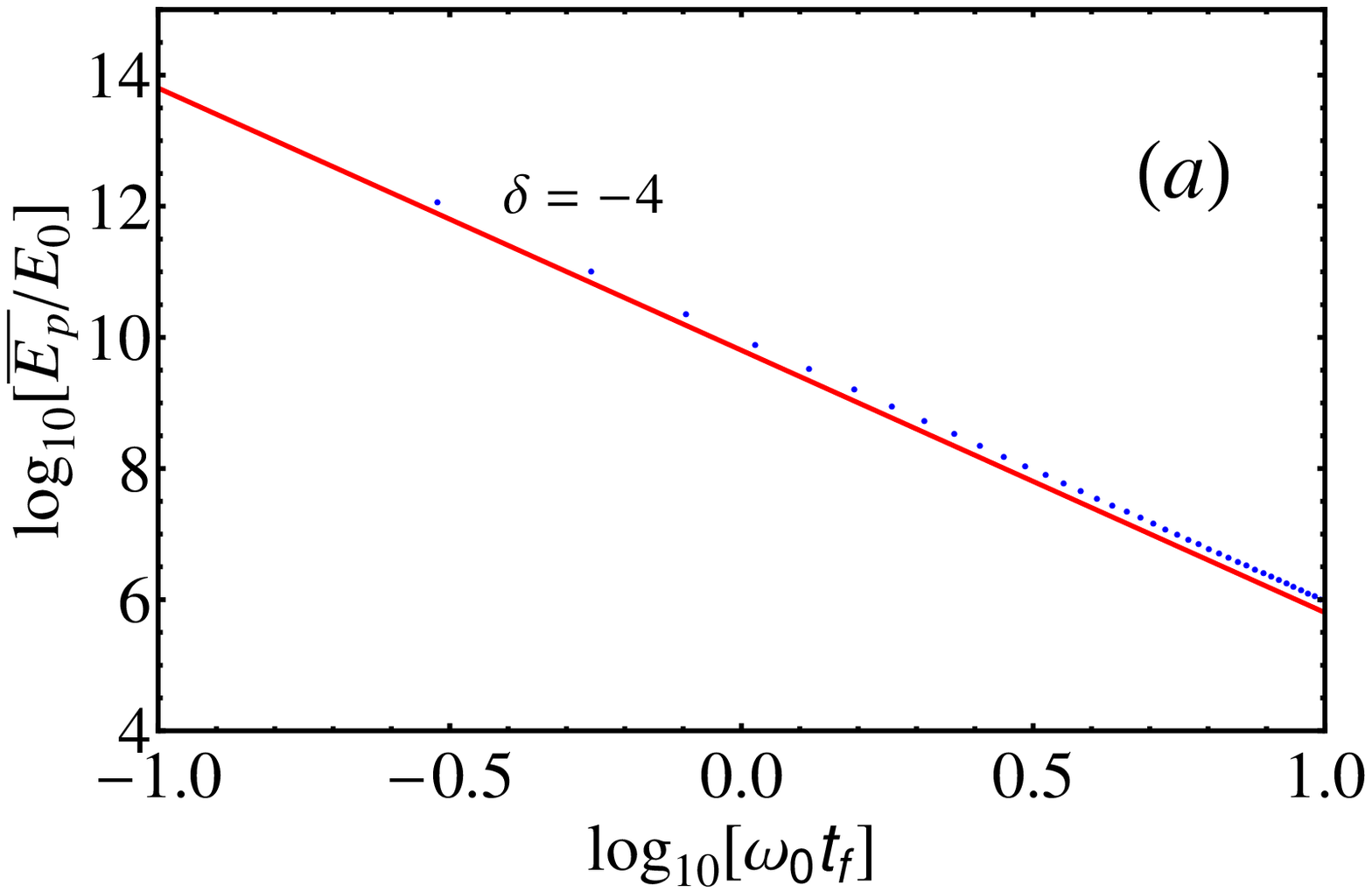}
\includegraphics[width=8cm]{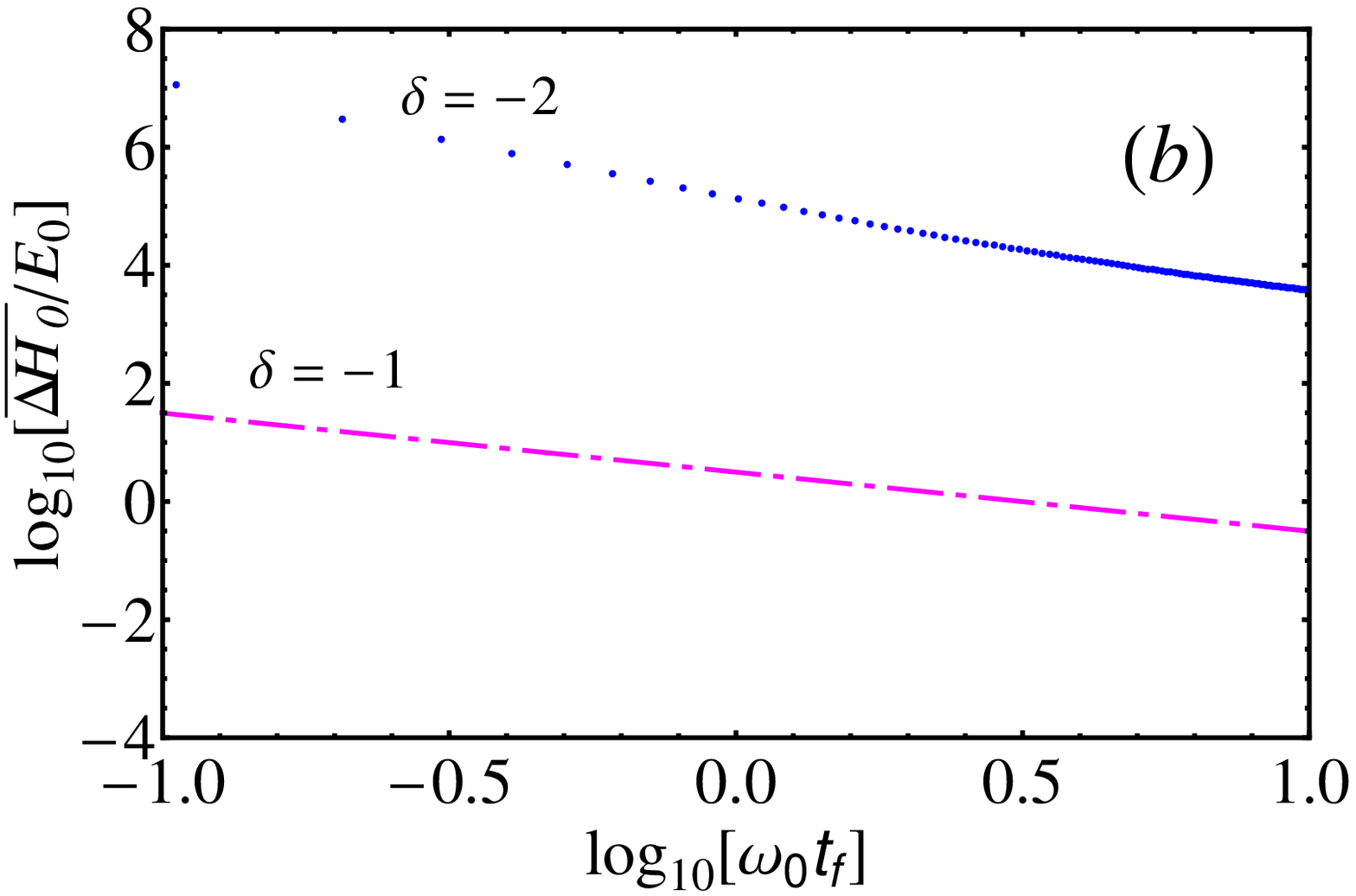}
\end{center}
\caption{(Color online) Dependences of time-averages energies with $t_f$. The $\delta$ is the asymptotic exponent of $t_f$. Parameters:   
$\omega_0=2 \pi\times 8$ Hz, $E_0=\hbar\omega_0/2$, $d=2.25$ mm, and rubidium-87 atoms that begin and end in the ground vibrational state.
(a) Bound (\ref{boundav}) (solid red line),
and time average of $E_P$ for a polynomial trajectory (dotted blue line).   
(b): AA Bound (dotted-dashed magenta line), 
and time average of $\Delta H$ for a polynomial trajectory (dotted blue line).   
}
\label{fig_tr}
\end{figure}
This bound describes the relevant dependences, as shown 
by numerical comparisons with actual time-averaged energies for polynomial trajectories, Eq. (\ref{pol}), see 
Fig. \ref{fig_tr}(a), and sets a rather strong $t_f^{-4}$ scaling,
compare this with the milder dependence on $t_f^{-2}$ of the time-averaged transient energy  in invariant-based inverse-engineered expansion processes \cite{energy}.  
As for the variance $(\Delta H)^2\equiv \la \psi_n|H^2|\psi_n\ra-\la\psi_n|H|\psi_n\ra^2$ for the $n$-th transport mode, it takes, after a somewhat lengthy calculation, a simple form,
\beq
(\Delta H)^2=2\hbar\omega_0(n+1/2)\left[\frac{1}{2}m
\omega_0^2
(q_c-q_0)^2+\frac{1}{2}m\dot{q}_c^2\right].
\eeq
Using again an Euler-Lagrange equation we find for 
its time average  the lower bound 
$\overline{(\Delta H)^2}> 12 \hbar(n+1/2)md^2/\omega_0 t_f^4$.
This does not establish a lower bound for the average of the 
standard deviation $\overline{\Delta H}$, but agrees with the  
scaling with $t_f$ that we observe numerically as $t_f\to 0$, $\overline{\Delta H}\propto t_f^{-2}$, see Fig. \ref{fig_tr}(b).  
This should be contrasted with the  Aharonov-Anandan (AA) relation \cite{AA} $\overline{\Delta H}\ge h/4t_f$, applied to transport among orthogonal states. (The general expression for ground-state to ground-state
transport allowing for non-orthogonal initial and final states \cite{AAPati,energy}  
is $\overline{\Delta H} t_f\ge \hbar\arccos[\exp(-m\omega_0d^2/4\hbar)]$, 
which tends to the result for orthogonal states when  $d>>(4\hbar/m\omega_0)^{1/2}$.)   
As it occurs for harmonic trap expansions \cite{energy},
while certainly correct as a bound, 
it does not describe the dependences found for 
the averaged standard deviation for fast
processes (small $t_f$).        
\subsection{Mean value theorem}
\begin{figure}[h!]
    \begin{center}
        \includegraphics[width=8cm]{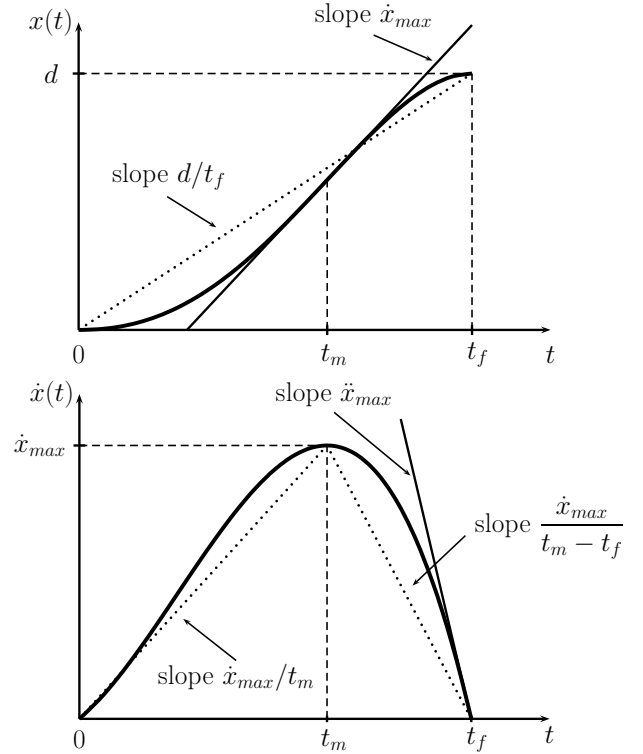}
    \end{center}
    \caption{Graphical representation of the lower bound of Eq.~(\ref{mvt1}) for the maximum velocity $\dot{x}_{max}$ (upper graph), and maximum acceleration $|\ddot{x}|_{max}$ of Eqs.~(\ref{mvt2}) and (\ref{mvt3}) (lower graph).}
    \label{fig_uplus}
\end{figure}
The mean value theorem (MVT) sets another useful bound since it applies to the instantaneous values rather than to a time average. 
The argument may be applied to $q_c$ in case ({\it i}) or to $q_0$ in case ({\it ii}), so we shall formulate it in terms of a generic $x$, assumed continuous in $[0, t_f]$,  
differentiable in $(0, t_f)$, and such that $x(0)=\dot{x}(0)=\dot{x}(t_f)=0$ and $x(t_f)=d$.
The maximum of its time 
derivative must be  
\beq
\dot{x}_{max}\ge d/t_f
\label{mvt1}
\eeq
at some point $t_{m}$ in $(0,t_f)$, see Fig. \ref{fig_uplus}.
We can now use that point to divide $[0,t_f]$ into two segments and apply 
the MVT again, now to the derivative.
In the first segment the derivative goes from 0 to 
$\dot{x}_{max}$ so  
\beq 
|\ddot{x}|_{max}\ge d/(t_f t_m).
\label{mvt2}
\eeq
Similarly in the second segment, 
\beq 
|\ddot{x}|_{max}\ge d/[t_f (t_f-t_m)],  
\label{mvt3}
\eeq
and, for the whole interval, $|\ddot{x}|_{max}\ge d/[t_f\, {\rm min}(t_m,t_f-t_m)]$.
Irrespective of the location of the point $t_m$,
${\rm min}(t_m,t_f-t_m)\le t_f/2$.  
We can thus set a lower bound for the (absolute value of) the maximum of the second derivative,  
\beq 
\label{bounda}
|\ddot{x}|_{max}\ge 2d/t_f^2. 
\eeq
When $x=q_c$, case ({\it{i}}), Eq. (\ref{bounda}) gives a bound for the instantaneous quasi-potential energy 
\beq
E_P\ge 2m\left(\frac{d}{\omega t_f^2}\right)^2.  
\eeq
In this case, however, we get a tighter lower bound 
directly from the time average (\ref{boundav}).   

With $x=q_0$, case ({\it ii}), we get a lower bound for the maximum trap acceleration. 
In particular the accelerations of compensating forces, typically limited by experimental constraints such as gradient magnetic fields or Stark electric fields,  should at the very least reach  the value $2d/t_f^2$.  
%
%
%
%
%
%
%
%
\section{Discussion\label{dis}}
We have applied the Lewis-Riesenfeld method \cite{LR} for quadratic-in-$p$ 
invariants \cite{DL}, combined with inverse engineering of trap trajectories, to design fast and faithful atomic transport.
The limitations have been quantified,  
and relations to other approaches that can be included in this framework have been pointed out. 
Another approach to accelerate adiabatic processes has been recently proposed by Berry \cite{Berry09} and can as well be applied formally to transport. The central idea is to construct a Hamiltonian that drives the system exactly along the adiabatic approximation defined for some reference time-dependent Hamiltonian $H_0(t)$, without transitions among the instantaneous eigenstates of $H_0$. For an arbitrary, rigid trap potential moving from $0$ to 
$d$ along a path $q_0(t)$, the eigenvectors of the instantaneous (reference) Hamiltonian $H_0=p^2/2m+U(q-q_0)$ are simply displaced from the original location, 
\beq
|n(t)\ra=e^{-i{p}q_0(t)/\hbar}|n(0)\ra.  
\eeq
The transitionless driving Hamiltonian $H_0+i\hbar\sum_n |\partial n\ra\la n|$ becomes in this case $H_0+{p}\dot{q}_0$, compare with Eq. (\ref{hamii}). 
Note the freedom to choose $H_0$. It could even be suppressed during transport: the simple Hamiltonian ${p}\dot{q}_0$ would also keep the same populations (of any $H_0$) without generating dynamical phases $e^{-iE_n t}$ for each eigenvalue. This is a rather intuitive result 
since the corresponding propagator is nothing but the displacement
operator $e^{-i{p}q_0(t)/\hbar}$. This approach provides thus a formal solution to fast and faithful transport, but the practical realization of a $\dot{q}_0{p}$ Hamiltonian term remains an open question.

 
With respect to the general framework embraced by Eqs. (\ref{Vinv}-\ref{last}),     
the studied cases termed ({\it{i}}) and ({\it{ii}}) in the main text are very relevant but not exhaustive. 
%
For $U=0$, the most general case occurs when transport is accompanied by expansions and contractions, so that 
the time dependence of $\omega(t)$ and $\rho(t)$ 
has to be considered if the invariant-based inverse engineering method is applied. 

For harmonic trap expansions 
or contractions in 
the gravity field ($q$ becomes a vertical coordinate), $F=-mg$ and the center of the trap suffers a time dependent translation $q_0=-2g/\omega^2$.
Again, $\alpha$ in Eq. (\ref{alphaeq}) may be interpreted
as a classical trajectory, now of a time dependent 
harmonic oscillator subjected to the gravity field, 
\beq
\ddot{q}_c+\omega^2(t)q_c=-g. 
\eeq
This is admittedly not a proper transport problem,  
but its formal treatment is the same, and has 
recently been implemented experimentally \cite{Nice}, also for 
Bose-Einstein condensates \cite{Nice2}.     

As for further extensions or open questions of the invariant approach,
one may investigate the use of more complex invariants, not restricted 
to being quadratic in $p$ \cite{LeeR}, in particular to tackle 
anharmonic transport. In the frame of quadratic-in-$p$ invariants, 
anharmonic traps can be dealt with by a compensating force (case ({\it{ii}})), but this force might be difficult to implement for large accelerations. If $F=0$, $\omega=0$, and $U(q-\alpha)\ne 0$ in Eq. (\ref{Vinv}), 
$\alpha$ should be the trap trajectory $q_0$, which is only consistent
with the auxiliary equation $\ddot{\alpha}=0$, see Eq. (\ref{alphaeq}),  for constant-velocity trajectories, incompatible with the boundary conditions (\ref{condq0}). A way out, to be explored, may be to use the invariants to implement minimization algorithms of the final vibrational excitation.    
   
Finally, further work will be devoted to understand and mitigate the effects of noise, for which the inversion method is intrinsically robust \cite{Calarco2009}, and of atom-atom interactions. 
The regime of Tonks-Girardeau gas can be treated along similar lines 
of the single particle case and for condensates,
scaling techniques may be used as in \cite{bec,Nice2}.       
\section*{Acknowledgments}
We thank T. Lahaye and M. Berry for discussions.   
We also acknowledge the kind hospitality of the Max Planck 
Institute for the Physics of Complex Systems in Dresden,  
funding by the Basque Government (Grant No. IT472-10),   
Ministerio de Ciencia e Innovaci\'on (Grant No. FIS2009-12773-C02-01), the Agence Nationale de la Recherche (Grant No. ANR-09-BLAN-0134-01) and the R\'{e}gion Midi-Pyr\'{e}n\'{e}es. E. T. and S. I. acknowledge financial support from the Basque Government (Grants No. BFI08.151 and BFI09.39). X. C. acknowledges support from Juan de la Cierva Programme and the National Natural Science Foundation of China (Grant No. 60806041).

\appendix

\section{Displacement unitary transformation}\label{B}
%
%
Consider the following (time dependent) 
position and momentum displacement unitary operator 
\beq
{\cal{U}}=e^{i{p}q_0(t)/\hbar}e^{-im\dot{q_0}(t){q}/\hbar}. 
\eeq
We could as well use variants of ${\cal{U}}$ with different orderings
without affecting the final result. 
Starting from the Schr\"odinger equation
\beq
i\hbar\partial_t|\psi\ra=H|\psi\ra,
\eeq
where, as in Eq. (\ref{hamii}),  
$H={p}^2/2m+U({q}-q_0)-m{q}\ddot{q}_0$, the corresponding equation for $|\Phi\ra={\cal U}|\psi\ra$ is   
\beqa
i\hbar \partial_t |\Phi\ra&=&{\cal U}H{\cal U}^\dagger|\Phi\ra+i\hbar(\partial_t
{\cal U})U^\dagger|\Phi\ra
\nonumber\\
&=&  
\left[\frac{{p}^2}{2m}+U({q})+\frac{m\dot{q}_0^2}{2}\right]|\Phi\ra.
\eeqa
%
Any stationary state in this ``trap frame'' will remain so 
in spite of the trap motion thanks to the compensating effect of the 
term $-m{q}\ddot{q}_0$ in $H$. 
%
%
%
%
%
%
%
%
%
%
%
%
\section{Perturbation theory analysis of the effect of anharmonicity\label{pera}}
In this Appendix we shall use perturbation theory to determine the effect 
of small anharmonicities using inverse or bang-bang trap trajectories.  
We start for concreteness 
from the ``cigar trap'' potential associated with a Gaussian beam with a moving focus, 
\beq
V(\Delta,r)=-{V_0}e^{-2r^2/w^2(\Delta)}\frac{1}{1+\frac{\Delta^2}{x_{R}^{2}}},
\eeq
where $r$ and $\Delta$ are radial and longitudinal coordinates, $\Delta=q-q_0(t)$, 
\beq
w(\Delta)=w_0\sqrt{1+\left(\frac{\Delta}{x_R}\right)^2}
\eeq
is the spot size, $x_R=\pi w_0^2/\lambda$ is the Rayleigh length,
and $w_0$ the waist. 
 
For a tight radial confinement we may ignore the radial coordinate 
and set $r=0$. The resulting longitudinal potential can be expanded around the minimum. Retaining the first correction to the harmonic term, and 
ignoring the constant, we split the Hamiltonian 
considering the quartic term as a perturbation,   
%
%
%
\beqa
H&=&H_0+V_1,
\nonumber \\
&=&\frac{p^2}{2m}+V_0\frac{[q-q_0(t)]^2}{x_{R}^{2}}-V_0\frac{[q-q_0(t)]^4}{x_{R}^{4}},
 \nonumber \\
V_1&=&-V_0\frac{[q-q_0(t)]^4}{x_{R}^{4}}, 
\eeqa
where $V_0=m\omega_0^2 x_R^2/2$.  
Using time dependent perturbation theory we calculate the overlap $\la\psi(t_f)|\tilde\psi(t_f)\ra$ between the state evolving with the
harmonic oscillator $|\psi(t)\ra$ and the perturbed state $|\tilde\psi(t)\ra$ at the final time $t=t_f$. $|\psi(t)\ra$ is chosen as the transport mode (\ref{psin}), with (\ref{con0}) satisfied.  

We approximate the perturbed state in first order as
\beqa
|\tilde\psi(t)\ra&=&U_0(t,0)|\tilde\psi(0)\ra
\nonumber \\
&-&\frac{i}{h}\int_{0}^{t}dt'U_{0}(t,t')V(t')U_0(t',0)|\tilde\psi(0)\ra, 
\eeqa
where 
\beq
U_0(t,0)=\exp\bigg(-\frac{i}{\hbar}\int_{0}^{t}dt'H_{0}(t')\bigg).
\eeq
so
\beqa
\label{e1}
&&\la\psi(t_f)|\tilde\psi(t_f)\ra=\la\psi(t_f)|U_{0}(t_f,0)|\tilde\psi(0)\ra 
\\
&-&\frac{i}{\hbar}\int_{0}^{t_f}dt'\la\psi(t_f)|U_{0}(t_f,t')V(t')U_0(t',0)|\tilde\psi(0)\ra.
\nonumber
\eeqa
At $t=0$ the initial state is also an eigenstate of the harmonic oscillator, $|\tilde\psi(0)\ra=|\psi(0)\ra$, so   
the first term in the right side of the previous equation is 1.
Using the transport modes in Eq. (\ref{psin}),
%
%
we calculate the bracket term in Eq. (\ref{e1}),
\beqa
\label{e2} &&\la\psi(t_f)|U_{0}(t_f,t')V(t')U_0(t',0)|\tilde\psi(0)\ra
\nonumber\\
&=&\frac{-V_{0}}{x_{R}^{4}}
\la\psi(t')|[q-q_{0}(t')]^4|\psi(t')\ra.
\eeqa
%
%
%
%
Using for $q_0$ and $q_c$ the functions for the bang-bang case 
described in Sec. \ref{bang-bang}, see Eqs. 
(\ref{Davs}) and (\ref{dife}), and 
performing the time integral we arrive, in first order, at  
\beq
\la\psi(t_f)|\tilde\psi(t_f)\ra=1-\frac{i}{\hbar}{\cal F},
\eeq
where the result for ${\cal F}$ is explicit but lengthy and not very illuminating so it is omitted here.  
%
%
It may be checked that when $t_f\rightarrow 0$ then 
${\cal F}\rightarrow 0$,
but this is not a very useful limit since the bang-bang procedure
will not work for times smaller than $4 \pi/\omega_0$. 
%
%
As a more useful case we take for $t_f$ the discrete times  $t_{f,N}=4\pi N/\omega_0$ with $N\in\mathbb{N}$ \cite{David}. Then 
${\cal F}$ takes the form
\beqa
\label{e5}
{\cal F}_{bb}&=&\frac{-2^{-(10+n)}(2n)!!}
{N^7m\pi^7n!\omega_0 x_{R}^2}
\\
&\times&\big\{1536N^8\hbar^2[1+2n(1+n)]\pi^8
\nonumber\\
&+&576d^2N^4\hbar m(1+2n)\pi^4\omega_0+35d^4m^2\omega_0^2\big\}.
\nonumber
\eeqa
In the limit $x_{R}\rightarrow\infty$, ${\cal F}_{bb}\rightarrow 0$. Increasing the waist, and keeping 
the other parameters constant, the  potential is more
harmonic. By contrast, as $x_{R}\rightarrow 0$, 
${\cal F}_{bb}\rightarrow -\infty$. Also, ${\cal F}_{bb}\rightarrow -\infty$ when 
$\omega_0\rightarrow 0$ and $\omega_0\rightarrow\infty$. 

If instead of the bang-bang functions we choose the inverse-engineered polynomial $q_c$ in Eq. (\ref{pol})  
and the corresponding $q_0$, ${\cal F}$ becomes,    
in the general case, i.e., for arbitrary parameters and in particular
an arbitrary $t_f$,   
\beqa
{\cal F}_{inv}&=&\frac{-2^{-(n+2)}V_{0}(2n)!!}{x_{R}^{4}n!}\bigg[\frac{1728000d^4}{1001t_{f}^{7}\omega_0^{8}}
\nonumber\\
&+&
\frac{1440d^2\hbar(1+2n)}{7mt_{f}^{3}\omega_0^{5}}+\frac{\hbar^2[3+6n(1+n)t_{f}]}{m^2\omega_0^2}\bigg].
\nonumber 
\eeqa
For the final times $t_{f,N}$, 
\beqa
\label{e6}
{\cal F}_{inv}&=&\frac{-2^{-(8+n)}3(2n)!!}
{1001N^7m\pi^7n!\omega_0 x_{R}^2}
\\
&\times&\big\{128128N^8\hbar^2[1+2n(1+n)]\pi^8
\nonumber\\
&+&34320d^2N^4\hbar m(1+2n)\pi^4\omega_0+1125d^4m^2\omega_0^2\big\}.
\nonumber
\eeqa
Comparing the factors in (\ref{e5}) and (\ref{e6}), we see that
$|{\cal F}_{inv}|< |{\cal F}_{bb}|$ for  $\omega_0>0$.


\begin{thebibliography}{999}     
%
\bibitem{Leibfried2002}
M. A. Rowe et al., Quant. Inf. Comp. \textbf{4}
257 (2002). 
%
\bibitem{ions} R. Reichle, D. Leibfried, R. B. Blakestad, J. Britton, J. D. Jost, E. Knill, C. Langer,
R. Ozeri, S. Seidelin, and D. J. Wineland, Fortschr. Phys. \textbf{54}, 666 (2006). 
%
\bibitem{MeschNature} Y. Miroschnychenko, W. Alt, 
I. Dotsenko, L. F\"orster, M. Khudaverdyan, D. Meschede, D. Schrader, and A. Rauschenbeutel, Nature \textbf{442}, 151 (2006). 
%
\bibitem{Ketterle2002}
T. L. Gustavson, A. P. Chikkatur, A. E. Leanhardt, A. G\"orlitz, S. Gupta, D. E. Pritchard, and W. Ketterle, Phys. Rev. Lett. \textbf{88}, 020401 (2002). 
%
\bibitem{Cornell2003}
H. J. Lewandowski, D. M. Harber, D. L. Whitaker, and E. A. Cornell, J. Low. Temp. Phys. \textbf{132}, 309 (2003).
%
\bibitem{Pritchard} M. J. Pritchard, A. S. Arnold, S. L. Cornish,
D. W. Hallwood, C. V. S. Pleasant, and I. G. Hughes, New J. Phys. \textbf{8}, 309 (2006). 
%
\bibitem{detector} G. R. Guth\"ohrlein, M. Keller, K. Hayasaka, W. Lange, and  H. Walther, Nature \textbf{414}, 49 (2001).
%
\bibitem{Calarco2000}
T. Calarco, E. A. Hinds, D. Jaksch, 
J. Schmiedmayer, J. I. Cirac, and P. Zoller, Phys. Rev. A \textbf{61}, 022304 (2000). 
%
\bibitem{Maleki} J. D. Prestage, R. L. Tjoelker, G. J. Dick, and L, Maleki, ``Improved Linear Ion Trap Package'',
Proc. 1993 IEEE Freq. Control Symposium, p. 144, June 1993.
%
\bibitem{Meschede} S. Kuhr, W. Alt, D. Schrader, M.M\"uller,
V. Gomer, and D. Meschede, Science \textbf{293}, 278 (2001). 
%
\bibitem{catcher1} S. Schmidt, J. G. Muga, and A. Ruschhaupt, Phys. Rev. A \textbf{80},
023406 (2009).
%
\bibitem{catcher2}X. Chen, J. G. Muga, A. del Campo, and A. Ruschhaupt, 
Phys. Rev. A \textbf{80}, 063421 (2009). 
%
\bibitem{HanschPRL2001} W. H\"ansel, J. Reichel, P. Hommelhoff, T. W. H\"ansch, 
Phys. Rev. Lett. \textbf{86}, 608 (2001).
%
\bibitem{HanschNature2001}W. H\"ansel, P. Hommelhoff, T. W. H\"ansch, and J. Reichel, 
Nature \textbf{413}, 498 (2001).
%
\bibitem{ocn}S. Kuhr, W. Alt, D. Schrader, I. Dotsenko, Y. Miroshnychenko, W. Rosenfeld, M. Khudaverdyan, V. Gomer,
A. Rauschenbeutel, and D. Meschede, Phys. Rev. Lett. \textbf{91}, 213002 (2003).
%
\bibitem{ocnsna}
D. Schrader, S. Kuhr, W. Alt, M. M\"uller, V. Gomer, and D. Meschede, Appl. Phys. B \textbf{73}, 819 (2001).
%
\bibitem{HanschPRA2001}M. Greiner, I. Bloch, T. W. H\"ansch, and 
T. Esslinger, Phys. Rev. A \textbf{63}, 031401 (2001).   
%
\bibitem{Lahaye} T. Lahaye, G. Reinaudi, Z. Wang, A. Couvert, 
and D. Gu\'ery-Odelin, Phys. Rev. A \textbf{74} 033622 (2006). 
%
\bibitem{David} A. Couvert, T. Kawalec, G. Reinaudi, and D. Gu\'ery-Odelin, Eur. Phys. Lett. \textbf{83}, 13001 (2008).
%
\bibitem{Denschlag}S. Schmid, G. Thalhammer, K. Winkler, F. Lang, and J. H. Denschlag, New J. Phys. \textbf{8}, 159 (2006).  
%
\bibitem{SK} G. Huber, T. Deuschle, W. Schnitzler, R. Reichle, K. Singer,
and F. Schmidt-Kaler, New J. Phys. \textbf{10}, 013004 (2008). 
%
\bibitem{Penning}
D. R. Crick, S. Donnellan, S. Ananthamurthy, R. C. Thompson, and D. M. Segal,
Rev. Sci. Instr. \textbf{81}, 013111 (2010). 
%
\bibitem{Wineland}
R. B. Blakestad, C. Ospelkaus, A. P. VanDevender, J.M. Amini, J. Britton, D. Leibfried, and D. J. Wineland, Phys. Rev. Lett. 
\textbf{102}, 153002 (2009).  
%
\bibitem{Calarco2009}
M. Murphy, L. Jiang, N. Khaneja, and T. Calarco, 
Phys. Rev. A \textbf{79}, 020301(R) (2009).
%
\bibitem{Nakamura2010}S. Masuda and K. Nakamura, Proc. R. Soc. A 
\textbf{466}, 1135 (2010).
%
\bibitem{Shan}D. Chen, H. Zhang, X. Xu, T. Li, and Y. Wang, Appl. Phys. Lett. \textbf{96}, 134103 (2010).
\bibitem{Chen}X. Chen, A. Ruschhaupt, S. Schmidt, A. del Campo, D. Gu\'ery-Odelin, and J. G. Muga, Phys. Rev. Lett. \textbf{104}, 063002 (2010).
%
\bibitem{LR}H. R. Lewis and W. B. Riesenfeld, 
J. Math. Phys. \textbf{10}, 1458 (1969).
%
\bibitem{Ermakov}V. P. Ermakov, Univ. Izv. Kiev. \textbf{20}, 1 (1880). 
%
\bibitem{LL} H. R. Lewis and P. G. Leach, J. Math. Phys. \textbf{23}, 
2371 (1982).
%
\bibitem{DL} A. K. Dhara and S. W. Lawande, 
J. Phys. A \textbf{17}, 2324 (1984).  
%
%
\bibitem{cg1} E. Narevicius et al., New J. Phys. \textbf{9}, 96 (2007).
%
\bibitem{cg2} M. Raizen, Science \textbf{324}, 1403 (2009)
%
\bibitem{sd1} H. L. Bethlem, G. Berden, G. Meijer, Phys. Rev. Lett. 
\textbf{83}, 1558 (1999).
%
\bibitem{sd2} S. Y. T. van de Meerakker, P. H. M. Smeets, N. Vanhaecke,
R. T. Jongma, G. Meijer, Phys. Rev. Lett. \textbf{94}, 023004 (2005).
%
\bibitem{sd3} B. C. Sawyer et al., Phys. Rev. Lett. \textbf{98}, 253002 (2007).
%
\bibitem{Fo} I. M. Gelfand and S. V. Fomin, Calculus of variations, Prentice Hall, New Jersey, 1963. 
%
\bibitem{energy} X. Chen and J. G. Muga,  arXiv:1009.5582.
%
\bibitem{AA} J. Anandan and Y. Aharonov, Phys. Rev. Lett.
\textbf{65}, 1697 (1990). 

\bibitem{AAPati} A. K. Pati, Phys. Lett. A \textbf{262}, 296 (1999). 
\bibitem{Berry09}M. V. Berry, J. Phys. A: Math. Theor. \textbf{42}, 365303
(2009).
\bibitem{Nice} J. F. Schaff, X.-L. Song, P. Vignolo, and G. Labeyrie, Phys. Rev. A \textbf{82}, 033430 (2010).
%
\bibitem{Nice2}J. F. Schaff, X. L. Song, P. Capuzzi, P. Vignolo, and G.  Labeyrie1, arXiv:1009.5868
\bibitem{LeeR}Y. H. Lee, J. Y.  Riu, J. Kor. Phys. Soc. \textbf{25}, 469 (1992). 
%
\bibitem{bec}J. G. Muga, X. Chen, A. Ruschhaupt and D Gu\'ery-Odelin, 
J. Phys. B: At. Mol. Opt. Phys. \textbf{42}, 241001 (2009).


   
%
%



\end{thebibliography}
\end{document}